\documentclass[12pt]{iopart}

\usepackage{color}

\usepackage{graphicx}

\begin{document}
\title[Constraining  non-minimally coupled tachyon fields by Noether symmetry ]
{Constraining  non-minimally coupled tachyon fields by Noether symmetry}

\author{Rudinei C. de Souza\dag\ and Gilberto M. Kremer\dag
\footnote[3]{To whom correspondence should be addressed (kremer@fisica.ufpr.br)}
}

\address{\dag\ Departamento de F\'\i sica, Universidade Federal do Paran\'a,
 Curitiba, Brazil}

\def\be{\begin{equation}}
\def\ee#1{\label{#1}\end{equation}}
\newcommand{\ben}{\begin{eqnarray}}
\newcommand{\n}{\nonumber}
\newcommand{\een}{\end{eqnarray}}
\newcommand{\lb}{\label}
\def\v{\varphi}

\begin{abstract}
A model for a homogeneous and isotropic Universe whose gravitational sources are a pressureless matter field and
a tachyon field non-minimally coupled to the gravitational field is analyzed. Noether symmetry is used to find
the expressions for the potential density and for the coupling function, and it is shown that both must be
exponential functions of the tachyon field. Two cosmological solutions are investigated: (i) for the early
Universe whose  only source of the gravitational field is a non-minimally coupled tachyon field  which behaves
as an inflaton and leads to an exponential accelerated expansion and (ii) for the late Universe whose
gravitational sources are a pressureless matter field and a non-minimally coupled tachyon field which plays the
role of dark energy and is the responsible of the decelerated-accelerated transition period.
\end{abstract}



\maketitle

\section{Introduction}

Presently it is well accepted by the scientific community that  the evolution of the Universe started with an
exponential accelerated expansion dominated by an entity called inflaton, followed by a decelerated period dominated
by matter fields and that presently the Universe has entered into a new accelerated period governed by a dark energy
field.

The search for models that could explain satisfactorily the inflationary period as well as the present accelerated
expansion is object of intense interest of the researchers in cosmology. Models which explains the accelerated expansion
of the Universe are based on scalar (see e.g. \cite{1})  or fermion (see e.g. \cite{2}) fields that are minimally or
non-minimally coupled to the gravitational field and which can simulate the inflaton in the primordial era and the dark
energy in the present period.

Within the context of scalar fields the models with tachyon fields minimally coupled to the gravitational field
received a considerable attention of the researchers. The tachyon field has its roots in string theory, but it
can be introduced in a simple manner as a generalization of the Lagrangian density of a relativistic particle,
i.e., $\mathcal{L}=-m \sqrt{1-\dot
q^2}\rightarrow\mathcal{L}_\v=-V(\varphi)\sqrt{1-\partial^\mu\varphi\partial_\mu\varphi}$, in the same way that
the quintessence could be considered as a generalization of the Lagrangian density for a non-relativistic
particle, namely, $\mathcal{L}=\dot
q^2/2-V(q)\rightarrow\mathcal{L}_\phi=\partial^\mu\phi\partial_\mu\phi/2-V(\phi)$. {In this work
it is discussed the role of the tachyon field within a cosmological framework and the search for any relationship  of the tachyon field to its root in string theory is not considered.}

In the references \cite{3}  the inflationary period of the Universe was investigated by using  tachyon fields
minimally coupled to the gravitational field for different self interaction potential densities, in the form of
power law, exponential and hyperbolic functions of the tachyon field. Tachyon fields with such kinds of
potential densities were also used in order to describe the present acceleration period of the Universe in the
references \cite{4} where the tachyon field behaves as dark energy. Furthermore, in the work \cite{5a} it was
also studied a tachyon field with an exponential potential density which can play the role of the inflaton and
dark energy. Although in the majority of those papers the tachyon field is minimally coupled to the
gravitational field, in reference \cite{5} it was investigated a non-minimally coupled tachyon field with
potential densities and coupling functions given by power law functions.

In all the above quoted works the potential densities were proposed in a ad-hoc way, with the aim the
determination of  desired cosmological solutions. The objective of the present work is to analyze a generic
model for a homogeneous and isotropic Universe whose gravitational sources are a matter field and a tachyon
field non-minimally coupled to the gravitational field. {The present model appears interesting
because it can describe in a reasonable way what is observed, since via Noether symmetry approach we can find a
unique form for the potential and coupling which describe the inflationary and the decelerated-accelerated
period.} The forms of the potential density and coupling function are determined from the Noether's symmetry
applied to the Lagrangian density which describes the model. Thus, in this paper Noether's symmetry approach
works as a criterion for the selection of the couplings and potentials instead of proposing their forms by an
ad-hoc way. By this proceeding we can restrict the possibilities of choice for the functions partially fixing
the potentials and couplings. Moreover, the existence of this symmetry guarantees conserved quantities, which
provide a motion constant that can help to integrate the field equations. Constraining scalar fields by Noether
symmetry is a subject of several papers in the literature (see e.g. \cite{6}).  In a recent paper \cite{6a} the
authors applied Noether symmetry for a fermion field non-minimally coupled with the gravitational field.

The evolution equations of the Universe follows from Einstein's field equations and  Klein-Gordon equation for
the coupled tachyon field which are solved for a given potential density and coupling function obtained from Noether
symmetry.
It is shown that a non-minimally coupled tachyon field could play the role of the inflaton describing the exponential
accelerated expansion in the early Universe and it could behaves as dark energy describing the  decelerated-accelerated
transition period of the late Universe.

The  organization of the work is the following: in the second section Einstein and Klein-Gordon equations are
derived from a point-like Lagrangian obtained from the action for a non-minimally coupled tachyon field and for
a Friedmann-Robertson-Walker metric. The purpose of the third section is the determination from Noether symmetry
of the possible forms of the self-interaction potential density and of the coupling function. The search of a
inflationary cosmological solution is the subject of the fourth section whereas in the fifth section the
decelerated-accelerated transition period is analyzed. Final remarks and conclusions in section sixth close this
work. From now on it is adopted the signature  $(+, -, -, -)$, the natural units $8 \pi G=c=\hbar=1$ whereas the
Ricci scalar in terms of the cosmic scale factor $a(t)$ is given by $$R=6\left(\frac{\ddot a}{a}+\frac{\dot a^2}{a^2}+\frac{k}{a^2}\right).$$

\section{Point-like Lagrangian and field equations}

Let $\varphi$ denote a rolling tachyon field with a Lagrangian density ${\cal L}_\v=
-V(\v)\sqrt{1-\partial_\mu\v\partial^\mu\v}$ where $V(\varphi)$ is the self-interaction potential density. The
action for a tachyon field non-minimally coupled to the gravitational field is written as
 \be
 S=\int\sqrt{-g}\ d^4x\ \bigg\{F(\varphi)R -
 V(\varphi)\sqrt{1-\partial_\mu\varphi\partial^\mu\varphi} \bigg\}+ S_m,
 \ee{1}
where $S_m$ is the action of a pressureless matter field, $R$ denotes Ricci scalar and $F(\varphi)$ represents
an arbitrary $C^2$ function of the tachyon field  which is related to the coupling of the
tachyon field with the gravitational field.

It is considered a homogeneous and isotropic Universe described by the Friedmann-Robertson-Walker metric
\be
ds^2=dt^2-a(t)^2\left[\frac{dr^2}{1-kr^2}+r^2(d\theta^2+\sin^2\theta d\phi^2)\right],
\ee{1a}
where $k=0,\pm1$.

 The point-like Lagrangian which follows from (\ref{1}) through a partial integration and by considering a homogeneous scalar field  $\varphi$=$\varphi(t)$, reads
\be
 \mathcal{L}=6a
 \dot{a}^2F+6a^2\dot{a}\dot{\varphi}{dF\over d\v} {-6kaF}+a^3V\sqrt{1-{\dot{\varphi}}^2}+\rho_m^0.
 \ee{2}
Above, $\rho_m^0$ is a constant value of the energy density of the matter field referred to an initial state and the dot denotes a derivative with respect to time.

The Klein-Gordon equation for the coupled tachyon field is obtained from the point-like Lagrangian (\ref{2}) through the use of Euler-Lagrange equation for $\v$, yielding
 \be
 \frac{\ddot{\varphi}}{1-\dot{\varphi}^2}+3H\dot{\varphi}+
 \frac{1}{V}\left[{dV\over d\v} -6\Bigg(\dot{H}+2H^2{+\frac{k}{a^2}}\Bigg)\sqrt{1-\dot{\varphi}^2}{dF\over d\v}\right]
 =0,
 \ee{3}
where $H=\dot a(t)/a(t)$ denotes the Hubble parameter.

Likewise, from Euler-Lagrange equation for the cosmic scale factor $a$ applied to the point-like Lagrangian (\ref{2}) one can obtain the acceleration equation, namely,
 \be
 \frac{\ddot{a}}{a}=-\frac{\rho+3p}{12F}.
 \ee{4}
In the above equation  $\rho=\rho_{\varphi}+\rho_m$ and
$p=p_{\varphi}$ are the energy density and the pressure of the sources of the gravitational field. The energy density and the pressure of the tachyon field are given by
 \ben\label{5}
 \rho_{\varphi}=\frac{V}{\sqrt{1-\dot{\varphi}^2}}-6H{dF\over d\v}\dot{\varphi},
 \\\label{6}
 p_{\varphi}=-V\sqrt{1-\dot{\varphi}^2}+2\left({dF\over d\v}\ddot{\varphi}+2H{dF\over d\v}\dot{\varphi}+{d^2F\over d\v^2}\dot{\varphi}^2\right),
 \een
respectively. Note that the matter field is considered as a pressureless fluid, i.e., $p_m=0$.

One can obtain Friedmann's equation
 \be
 H^2=\frac{\rho}{6F}{-\frac{k}{a^2}},
 \ee{7}
by imposing that the energy function associated to the point-like Lagrangian vanishes, i.e.,
 \be
 E_\mathcal{L}=\frac{\partial \mathcal{L}}{\partial \dot{a}}
 \dot{a}+\frac{\partial
 \mathcal{L}}{\partial\dot{\varphi}}\dot{\varphi}-\mathcal{L}\equiv0.
 \ee{8}
The above expression is another independent field equation when one considers a Friedmann-Robertson-Walker metric and a homogeneous scalar field in the action (\ref{1}).

\section{Constraints from Noether symmetry}

Let $\textbf{X}$ be the following infinitesimal generator of symmetry
 \be
 \textbf{X}=\alpha\frac{\partial}{\partial
 a}+\beta\frac{\partial}{\partial \varphi}+\Bigg(\frac{\partial
 \alpha}{\partial a}\dot{a}+\frac{\partial \alpha}{\partial
 \varphi}\dot{\varphi}\Bigg)\frac{\partial}{\partial
 \dot{a}}+\Bigg(\frac{\partial \beta}{\partial
 a}\dot{a}+\frac{\partial \beta}{\partial
 \varphi}\dot{\varphi}\Bigg)\frac{\partial}{\partial \dot{\varphi}},
 \ee{9}
where $\alpha$ and $\beta$ are only function of  $(a, \varphi)$. Furthermore, let $L_\textbf{x}$ denote Lie's derivative of the point-like Lagrangian $\mathcal{L}$ with respect to the vector $\textbf{X}$ which is defined in the tangent space.

Noether's symmetry is satisfied by the condition $L_\textbf{x}\mathcal{L}=0$ , i.e., $\textbf{X}\,\mathcal{L}=0$ which implies
 \ben\nonumber
 \alpha\left(6\dot{a}^2F+12a\dot{a}\dot{\varphi}{dF\over d\v}+3a^2V\sqrt{1-{\dot{\varphi}}^2}{-6kF}\right)+\beta\left(6a
 \dot{a}^2{dF\over d\v}+6a^2\dot{a}\dot{\varphi}{d^2F\over d\v^2}\right.
 \\\nonumber\label{10}\left.+a^3{dV\over d\v}\sqrt{1-{\dot{\varphi}}^2}{-6ka\frac{dF}{d\v}}\right)
 +\left(\frac{\partial \alpha}{\partial a}\dot{a}+\frac{\partial
 \alpha}{\partial \varphi}\dot{\varphi}\right)\left(12a
 \dot{a}F+6a^2\dot{\varphi}{dF\over d\v}\right)
 \\+\left(\frac{\partial
 \beta}{\partial a}\dot{a}+\frac{\partial \beta}{\partial
 \varphi}\dot{\varphi}\right)\left(6a^2\dot{a}{dF\over d\v}-\frac{a^3\dot{\varphi}V}{\sqrt{1-{\dot{\varphi}}^2}}\right)=0.
 \een

{After some rearrangements the equation (\ref{10}) depends explicitly on $\dot a$, $\dot\v$,
their powers and combinations with $\sqrt{1-\dot\varphi^2}$}, hence their
coefficients must vanish, yielding
 \ben\label{11}
 \left(\alpha+2a\frac{\partial \alpha}{\partial
 a}\right)F+\left(\beta+a\frac{\partial \beta}{\partial
 a}\right)a{dF\over d\v}=0,
 \\\label{12}
 2F\frac{\partial \alpha}{\partial
 \varphi}+\left(2\alpha+a\frac{\partial \alpha}{\partial
 a}+a\frac{\partial \beta}{\partial \varphi}\right){dF\over d\v}+a\beta
 {d^2F\over d\v^2}=0,
 \\\label{13}
 \frac{\partial \alpha}{\partial \varphi}{dF\over d\v}=0, \qquad
 \frac{\partial \beta}{\partial a}=0, \qquad \frac{\partial
 \beta}{\partial \varphi}=0,
 \\\label{14}
 3\alpha V+a\beta {dV\over d\v}=0,\\
 {\alpha F+a\beta{dF\over d\v}=0} \label{14.1}.
 \een
{The equations (\ref{11}) - (\ref{14.1}) come by imposing that the coefficients of $\dot a^2$,
$\dot a\dot\varphi$, $\dot\varphi^2$, $\dot a\dot\varphi/\sqrt{1-\dot\varphi^2}$,
$\dot\varphi^2/\sqrt{1-\dot\varphi^2}$ and $\sqrt{1-\dot\varphi^2}$ vanish. Moreover, it was supposed that $\dot
a$ and $\dot\varphi$ do not vanish during the time evolution of the Universe described by this model and that
the restriction $\dot \varphi^2\neq1$ holds. Observe that the simplifications related to $a$ -- which were done
to obtain the equations (\ref{11}) - (\ref{14.1}) -- are possible because it was considered that $a\neq0$.}

The analysis of the system of equations (\ref{11}) through (\ref{14.1}) proceeds as follows. First, from
equations (\ref{13})$_2$ and (\ref{13})$_3$ one infers that $\beta\equiv\beta_0$ must be constant. Next, from
equation (\ref{13})$_1$ two cases must be considered, namely, $dF/d\v=0$ and $dF/d\v\neq0$.

{The equation (\ref{14.1}) cannot be satisfied simultaneously with the rest of the system for the
case $dF/d\v\neq0$. When we take the case $dF/d\v=0$, the system is solved if $F=0$, which excludes the
gravitational field from the action (\ref{1}). This means that if the Noether symmetry must be satisfied the case $k\neq0$ is ruled out. But when one considers $k=0$
-- a flat Universe -- for the point-like Lagrangian (\ref{2}) the condition $L_\textbf{x}\mathcal{L}=0$
furnishes the system (\ref{11}) - (\ref{14.1}) without the equation (\ref{14.1}) and the system can present an
interesting solution. }
{Hence, the following analysis will be done for a homogeneous, isotropic and spatially flat Universe
and   it will be  considered in  all remaining equations that $k=0$. Below the two cases ($dF/d\v\neq0$ and $dF/d\v=0$) will be examined separately for the unique permitted situation by the Noether symmetry i.e., for $k=0$.}

\noindent \textbf{(a)} \emph{First case} $dF/d\v=0$

From the system of equations (\ref{11}) and (\ref{12}) it follows that $\alpha$ do not depend on $\v$ and is
proportional to $1/\sqrt{a}$. Further  one concludes that the only possibility to fulfill equation (\ref{14}) is
by taking $V=0$. This case is not interesting because the energy density and the pressure of tachyon field
vanish identically.

\noindent \textbf{(b)} \emph{Second case} $dF/d\v\neq0$

For $dF/d\v\neq0$ it follows from (\ref{13})$_1$ that  $\alpha$ is only a function of $a$ and the only possibility for equation (\ref{14}) to be satisfied is that
 \be
 \alpha=\alpha_0 a\qquad\hbox{and}\qquad V=\lambda\exp(-\xi\varphi),
 \ee{15}
where $\alpha_0$, $\lambda$ and $\xi=3\alpha_0/\beta_0$ are constants. Finally, from the system of equations (\ref{11}) and (\ref{12}) one can get that the coupling function reads
 \be
 F=\gamma\exp(-\xi\varphi),
 \ee{16}
where $\gamma$ is another constant.

The constant of motion associated with the Noether symmetry is
 \be
 \Sigma_0=\alpha\frac{\partial \mathcal{L}}{\partial
 \dot{a}}+\beta\frac{\partial \mathcal{L}}{\partial \dot{\varphi}},
 \ee{19}
which for the symmetry found leads to
 \be
 \frac{\Sigma_0}{6\gamma
 \beta_0}=-\left[\left(\frac{\kappa}{\sqrt{1-\dot{\varphi}^2}}+\frac{\xi^2}{3}\right)
 \dot{\varphi}+\frac{\xi}{3}H\right]a^3\exp(-\xi\varphi),
 \ee{20}
where $\kappa=\lambda/6\gamma$. This is the form of the constant of motion associated with the Noether symmetry
which was found for the dynamics described by the general action (\ref{1}). {Such a constant of
motion is a new conserved quantity which does not have any relation to some known a priori quantity.
}

Note that the non-interacting and pressureless matter field is proportional to $a^3$ and it appears as a
constant in the point-like Lagrangian (\ref{2}). Furthermore, it does not have any influence on the equations
(11)-(14) which result from the symmetry condition. Then the constant of motion associated with the symmetry is
the same for the cases with and without a matter field.

\section{Inflationary period}

In the inflationary period the role played by a matter field is negligible, so that one can rid of it by letting $\rho_m^0=0$. In this case one has to solve the system of coupled differential equations that follow from equations (\ref{3}) and (\ref{7}), namely,
 \ben\label{17}
 \frac{\ddot{\varphi}}{1-\dot{\varphi}^2}+3H\dot{\varphi}+\frac{\xi}{\kappa}(\dot{H}+2H^2)
 \sqrt{1-\dot{\varphi}^2}-\xi=0,
 \\\label{18}
 H^2-\xi H\dot{\varphi}-\frac{\kappa}{\sqrt{1-\dot{\varphi}^2}}=0,
 \een
by using equations (\ref{5}), (\ref{15}) and (\ref{16}).

For the determination of an analytical solution of the coupled system of differential equations (\ref{17}) and
(\ref{18}) one makes use of the constant of motion associated with Noether symmetry (\ref{20}), which after
differentiation with respect to $\v$ becomes an ordinary differential equation for $\v$ that can be written as
 \be
 \left(\frac{3\kappa}{\xi^2\sqrt{1-\dot{\varphi}^2}}+\frac{3}{2}\right)\dot{\varphi}
 \pm\sqrt{\left(\frac{\kappa}{\xi^2\sqrt{1-\dot{\varphi}^2}}+\frac{\dot{\varphi}^2}{4}\right)}=0,
 \ee{21}
thanks to (\ref{18}) and $\partial H/\partial\v=0$.

By considering $\xi=-\imath\xi_0$, where $\xi_0$ is a  real constant, the solution of the differential equation (\ref{21})  is given by
 \be
 \varphi(t)=\imath(k_1t+k_2),
 \ee{22}
where $k_1$ and  $k_2$ are  real constants. This solution implies the following  relationships for the constants
 \be
 \kappa=\frac{\lambda}{6\gamma}=\frac{1}{18}\left(\frac{\xi_0}{k_1}\right)^2\sqrt{1+k_1^2}
 \left(1+\sqrt{1+18k_1^2+9k_1^4}+9k_1^2\right).
 \ee{23}

Now the substitution of the time evolution of the tachyon field (\ref{22}) into the expressions for the potential density (\ref{15}) and coupling (\ref{16}) leads to
 \be
 V(t)=\lambda\exp[-\xi_0(k_1t+k_2)],
 \qquad
 F(t)=\gamma\exp[-\xi_0(k_1t+k_2)],
 \ee{24}
respectively. By requiring that the potential density must decay with time, one infers from  equation
(\ref{24})$_1$ that  $\xi_0k_1>0$.

Finally, one can obtain from equation (\ref{18}) through integration the time evolution of the cosmic scale
factor, namely,
 \be
 a(t)=\exp[K(t-t_0)],\quad\hbox{where}\quad
 K=\sqrt{\frac{\kappa}{\sqrt{(1+k_1^2)}}+\frac{\xi_0^2k_1^2}{4}}+\frac{\xi_0k_1}{2}.
 \ee{25}
Hence, this solution describes an inflationary period, where the cosmic scale factor increases exponentially with time.

The time evolution of the energy density and pressure of the tachyon field can be obtained from equations
(\ref{5}), (\ref{6}), (\ref{22}) and (\ref{24}), yielding
 \ben\label{26}
 \rho_{\varphi}(t)=\left(\frac{\lambda}{\sqrt{1+k_1^2}}+6\xi_0\gamma
 k_1K\right)\exp\left[-\xi_0(k_1t+k_2)\right],
 \\\label{27}
 p_{\varphi}(t)
 =-\frac{\lambda\sqrt{1+k_1^2}+2\xi_0\gamma
 k_1(2K-\xi_0k_1)}{\frac{\lambda}{\sqrt{1+k_1^2}}+6\xi_0\gamma
 k_1K}\rho_{\varphi}(t).
 \een
From the above equations one concludes that (i) the  pressure is proportional to energy density and is negative and (ii) the inflationary period comes to an end at a finite time since the energy density and pressure of the tachyon field tend to zero after that time.

\section{Decelerated-accelerated transition period}

The analysis of the case where the sources of the gravitational field are the tachyon and the pressureless matter fields proceeds by changing the variable and introducing the red-shift $z$ through the relationships
 \be
 z=\frac{1}{a}-1,\qquad \frac{d}{dt}=-H(1+z)\frac{d}{dz}.
 \ee{28}
In terms of the red-shift the Klein-Gordon (\ref{3}) and acceleration (\ref{4}) equations become
\ben
H^2(1 + z)^2\varphi''+\left[H+H'(1 + z)\right]H(1 + z)\varphi'=\left[1-H^2(1+
z)^2\varphi'^2\right]
\nonumber\\\label{29}
\times\left\{3H^2(1 + z)\varphi'+\frac{\xi}{\kappa}H\left[H'(1 +
z)-2H\right]\sqrt{1-H^2(1+z)^2\varphi'^2}+\xi\right\},
\\\label{30}
4\gamma
HH'\exp(-\xi\varphi)(1+z)=\rho_m+\rho_{\varphi}+p_{\varphi},
\een
where the energy densities of the matter and tachyon fields and the pressure of the tachyon field read
\ben\label{31}
\rho_m(z)=\rho_m^0(1+z)^3,
\\\label{32}
\rho_{\varphi}(z)=\exp(-\xi\varphi)\left\{\frac{\lambda}{\sqrt{1-H^2(1+z)^2\varphi'^2}}-6\xi
\gamma H^2(1+z)\varphi'\right\},
\\
p_{\varphi}(z)=-\exp(-\xi\varphi)\left\{\lambda\sqrt{1-H^2(1+z)^2\varphi'^2}+2\xi\gamma
\left[H^2(1+z)^2\varphi''\right.\right.
\nonumber\\\label{33}\left.\left.
\qquad\qquad+\Big(H'(1 + z)-H\Big)H(1
+ z)\varphi'-\xi H^2(1+z)^2\varphi'^2\right]\right\},
\een
respectively. In the above equations the prime denotes the derivative with respect to the red-shift $z$.

The search for exact solutions of the coupled system of differential equations (\ref{29}) and (\ref{30}) is a very hard job and a numerical solution of the system of equations will be analyzed afterwards.

For the determination of numerical solutions of the system of differential equations (\ref{29}) and (\ref{30}) one has to specify initial conditions for  $\varphi(z)$, $\varphi'(z)$ and  $H(z)$ at present time, i.e., at $z=0$. As usual  the energy densities are replaced by the density parameters $\Omega_\v(z)=\rho_\v(z)/\rho(z)$ and $\Omega_m(z)=\rho_m(z)/\rho(z)$ where $\rho(z)=\rho_\v(z)+\rho_m(z)$ is the total energy density of the sources of the gravitational field. Furthermore, one introduces the dimensionless quantities $H_*(z)=H(z)/\sqrt{\rho(0)}$, $\v_*(z)=\sqrt{\rho(0)}\v(z)$, $\xi_*=\xi/\sqrt{\rho(0)}$ and $\lambda_*=\lambda/{\rho(0)}.$ Here the  present values of the energy densities  adopted are $\Omega_\v(0)=0.72$ and $\Omega_m(0)=0.28$ (see e.g. reference \cite{Peebles}).

One expects that the present value of the coupling is $F=c^3/16\pi G$ or $F=1/2$ in natural units. Hence, it
follows from (\ref{16}) that $\varphi_*(0)=\ln(2\gamma)^{1/\xi_*}$. Besides, since in the late time the Universe
is accelerating, the negative pressure of the tachyon field must be the responsible for its acceleration. To
guarantee this behavior one imposes the condition $\dot\varphi^2(0)\ll1$. It follows from equation (\ref{7})
that $\rho_{\varphi}(0)\approx V(0)$, so that $\varphi_*(0)=\ln(\lambda_*/0.72)^{1/\xi_*}$, which together with
the previous expression leads to the relationship $\gamma=\lambda_*/1.44$.

From the above considerations among the three free parameters $\xi_*$, $\lambda_*$ and $\gamma$ only two are linearly independent. To find the numerical solutions of the system of differential equations it was chosen $\lambda_*=1$ and three values for the parameter which is related with the strength of the coupling, namely, $\xi_*=0.05$; 0.10; 0.20. Furthermore, the desired conditions at $z=0$ -- by taking into account that $\dot\varphi^2(0)\ll1$ -- are
\be
H_*(0)=\sqrt{1/3},\qquad \varphi_*(0)=-\ln(0.72)^{1/\xi_*},\qquad \v_*'(0)=10^{-3}.
\ee{34}

\begin{figure}
 \begin{center}
 \vskip0.5cm
 \includegraphics[height=5cm,width=7cm]{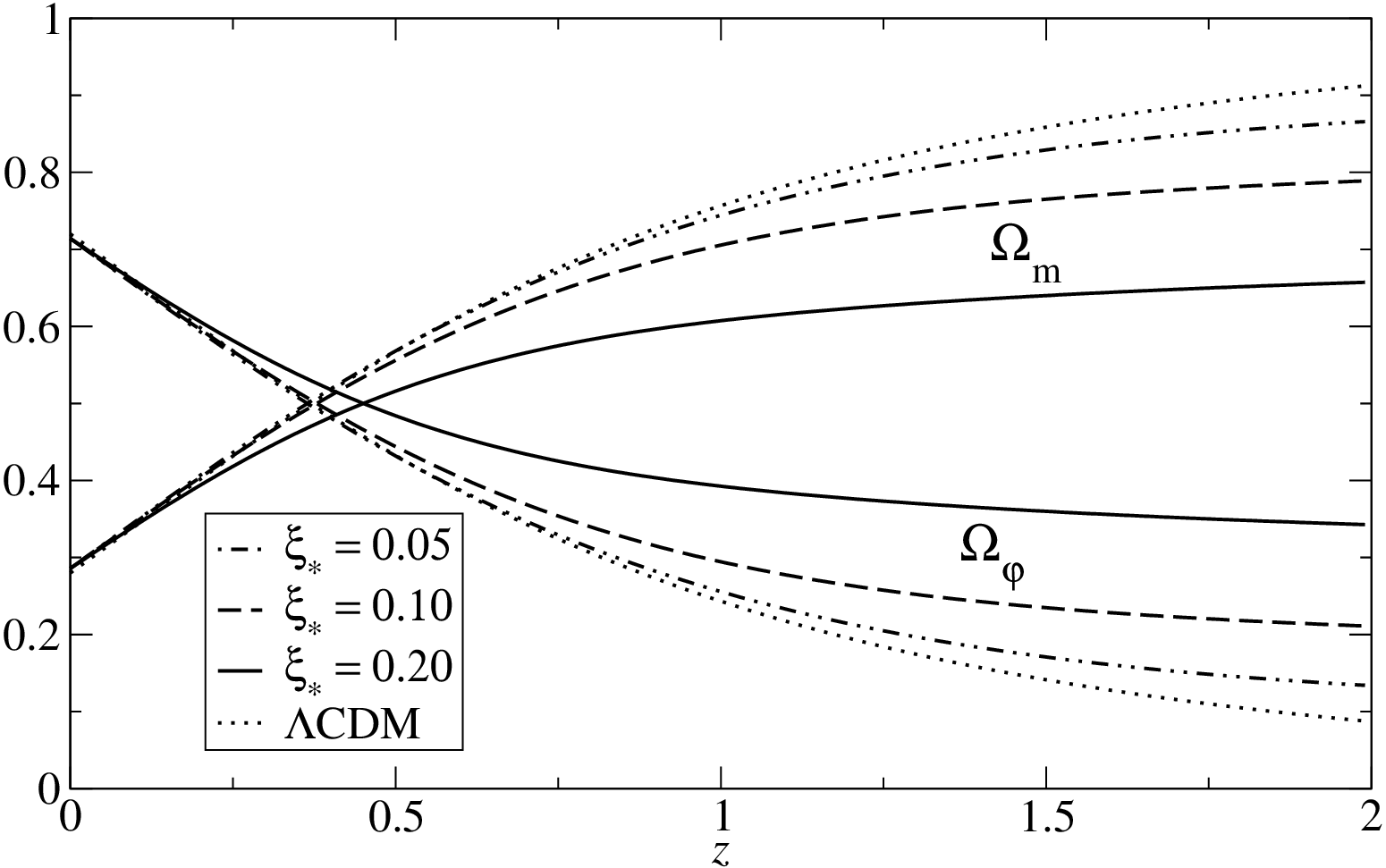}\hskip0.7cm
  \includegraphics[height=5cm,width=7cm]{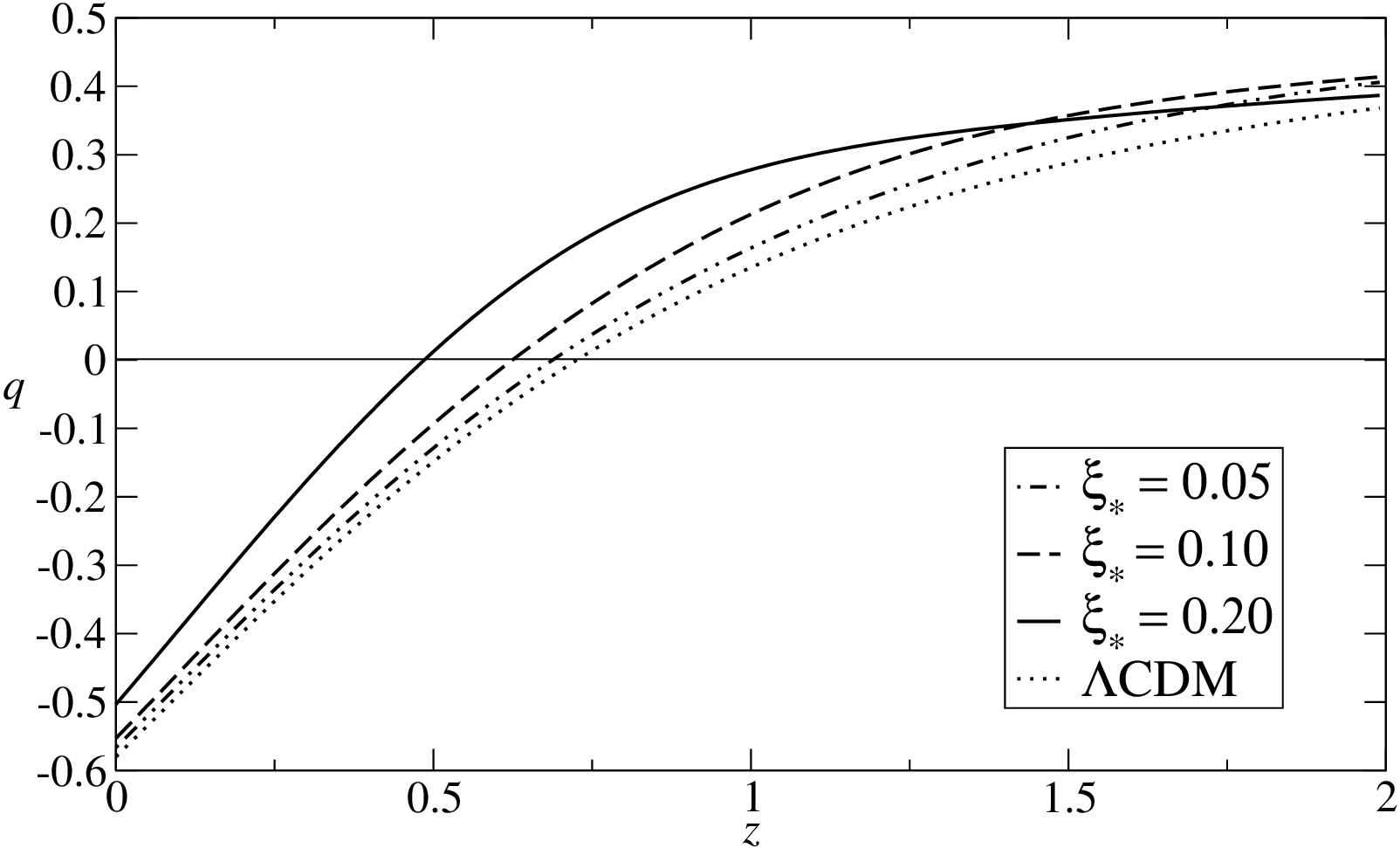}
  \caption{Left frame: density parameters of matter and tachyon fields as functions of the red-shift. Right frame: deceleration parameter as function of the red-shift.}
 \end{center}
 \end{figure}

In the left frame of figure 1 the density parameters of the matter and tachyon fields are plotted as a function of the red-shift in the range $0\leq z\leq2$. One observes from this figure that the coupling parameter $\xi_*$ has influence on the decay of the density parameter of the tachyon field and the corresponding increase of the matter field, being more efficient according as the value of $\xi_*$ decreases.
This behavior can be explained by noting that the increase of the coupling parameter $\xi_*$ increases the energy transfer from the tachyon field to the gravitational field.
Furthermore, as was expected, the limiting case $\xi_*\rightarrow0$ tends to the $\Lambda$CDM model.

The deceleration parameter  $q=1/2+3p/2\rho$ is plotted as function of the red-shift $0\leq z\leq2$ in the right frame of figure 1, whereas in table 1 it is compared the  values taken from the curves with the predicted values of the present  deceleration parameter $q(0)$ and the transition red-shift $z_t$ where the decelerated period turns over an accelerated period. From table 1 one infers that all values are almost within the uncertainty of the predicted values. Furthermore, by increasing the value of the coupling the red-shift transition $z_t$ decreases whereas the deceleration parameter $q(0)$ increases. It is noteworthy to call attention that for high values of the red-shift the coupling is not important, since the deceleration parameters for different values of coupling parameter $\xi_*$ tend to a common value.
\begin{table}
\centering
\begin{tabular}{|c|c|c|c|c|c|} \hline
 & $\xi_*=0.20$& $\xi_*=0.10$ & $\xi_*=0.05$&$\Lambda$CDM& predicted values \\
\hline
$q(0)$&-0.50&-0.55&-0.57 & -0.58&-0.74 $\pm$ 0.18 (from \cite{Virey})\\
\hline
$z_t$&0.49&0.62& 0.69& 0.73&0.46 $\pm$ 0.13 (from \cite{Riess})\\
\hline\end{tabular}
\caption{Values of the deceleration parameter $q(0)$ and of the red-shift transition $z_t$
for different values of the coupling parameter $\xi_*$.}
\end{table}

It is also interesting to analyze the evolution with the red-shift of the coupling function $F(z)$ for different values of the coupling parameter $\xi_*$. The plot of $F(z)$ versus $z$ is given in the left frame of figure 2 where one can infer that the coupling parameter $\xi_*$ has a prominent role on the behavior of $F$. Indeed, the variation of the coupling function $F$ is less accentuated for small values of the coupling constant $\xi_*$. This behavior was expected due to the expression of $F$ given by equation (\ref{21}). Another point is that for different values of the coupling parameter $\xi_*$ the coupling function tends to an asymptotic value by increasing values of the red-shift. The behavior of the potential density $V$ with the red-shift is similar to the coupling function $F$, since they have the same exponential dependence on $\v$.

In the right frame of figure 2 it is plotted the ratio  of the pressure and energy density of the tachyon field $\omega_\v=p_\v/\rho_\v$ as function of the red-shift $z$. One observes from this figure that $\omega_\v\rightarrow-1$ when $z\rightarrow0$ for small values of the coupling parameter $\xi_*$, i.e., for small values of $\xi_*$ the tachyon field  approaches the $\Lambda$CDM model. For high values of the red-shift $\omega_\v \rightarrow 0$, i.e., the tachyon field tends to a pressureless matter field.

\begin{figure}
 \begin{center}
 \vskip0.5cm
 \includegraphics[height=5cm,width=7cm]{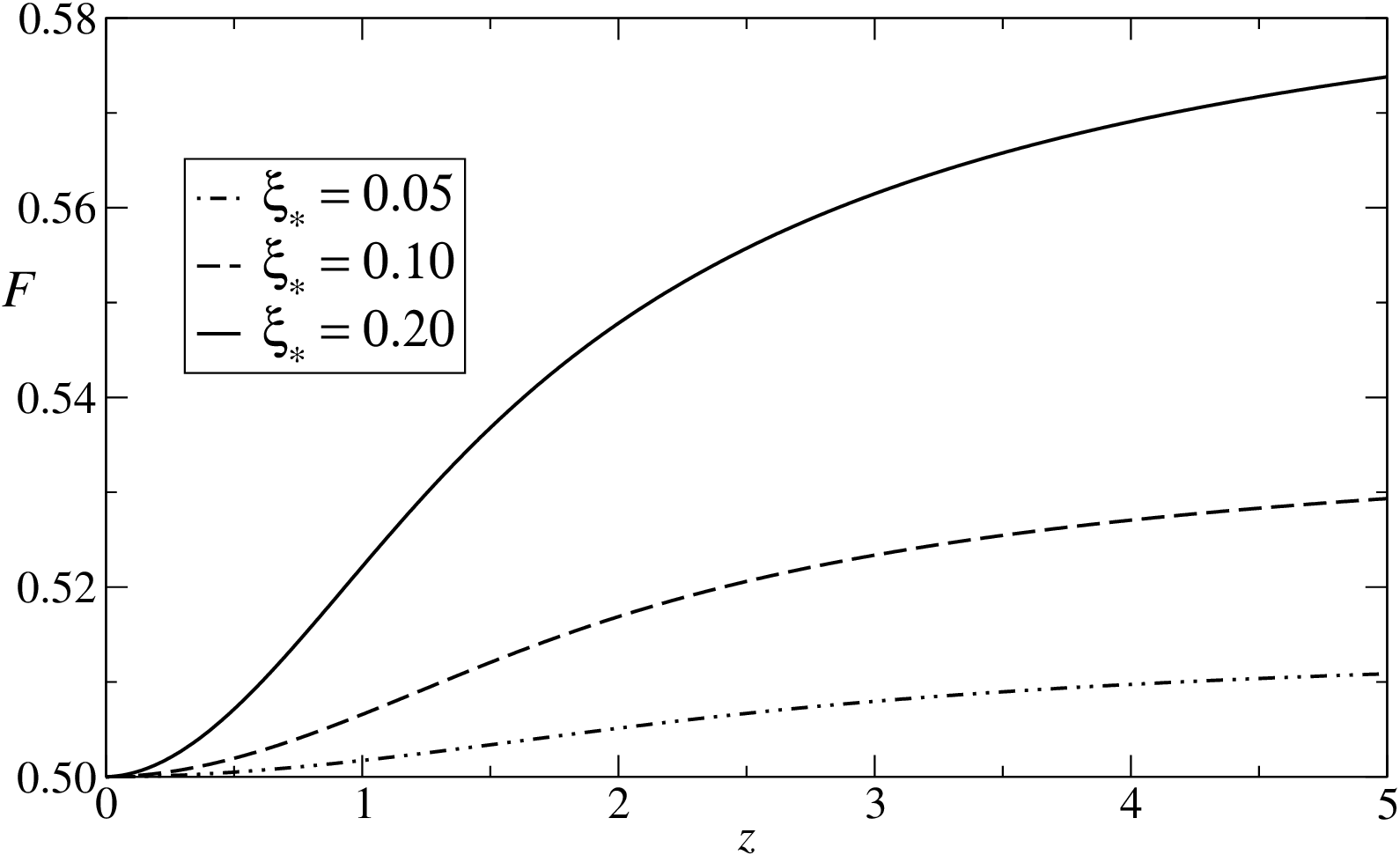}\hskip0.5cm
 \includegraphics[height=5cm,width=7cm]{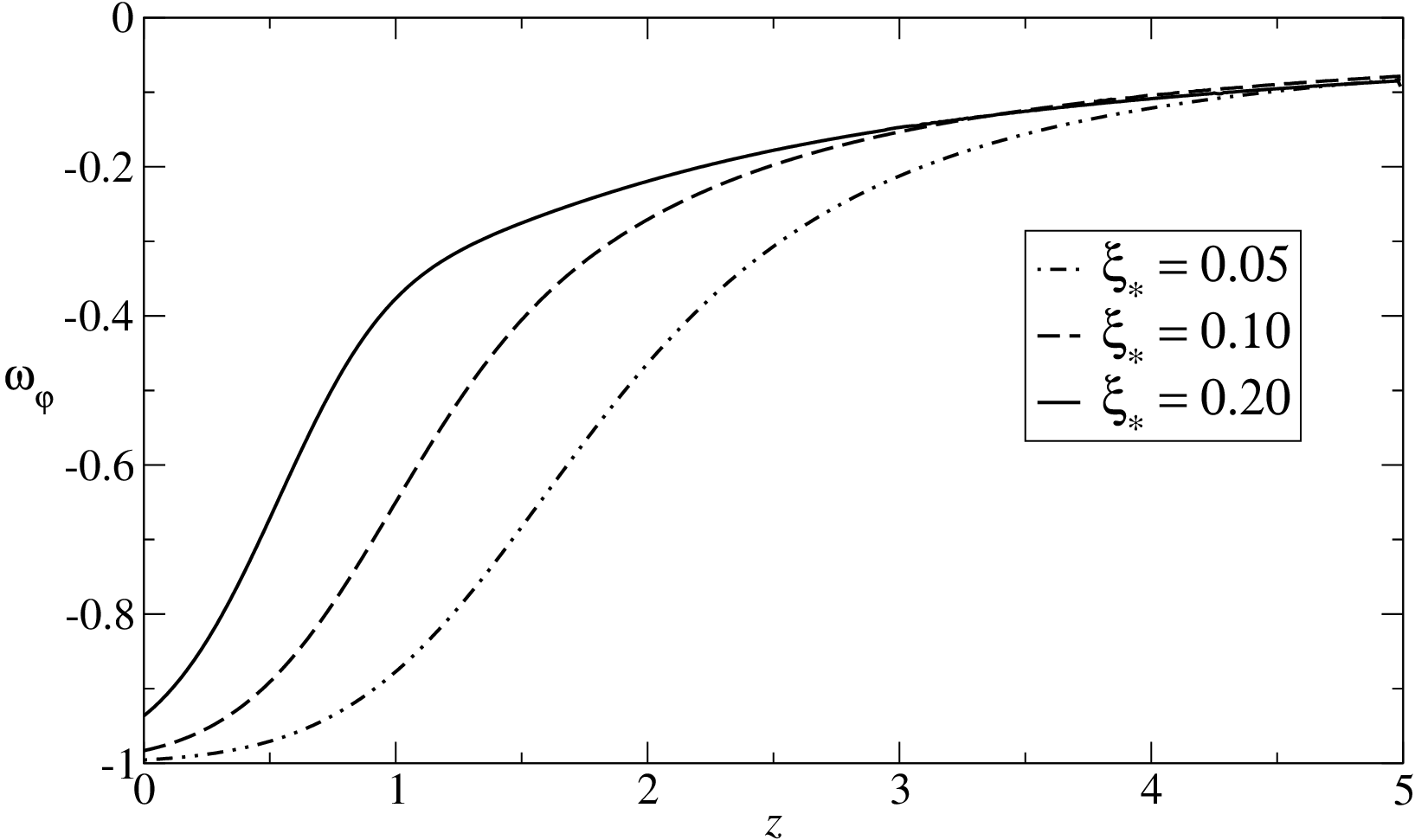}
  \caption{Left frame: coupling function $F$ versus the red-shift $z$. Right frame: ratio of  pressure and energy density of the tachyon field $\omega_\v$ versus the red-shift $z$.  }
 \end{center}
 \end{figure}

The difference of the apparent and absolute magnitude of a source is defined in terms of the luminosity distance and given by
 \be
 \mu_0=m-M=25+5\log\left[(1+z)\int_0^z\frac{dz'}{H(z')}\right].
 \ee{35}
It is plotted in figure 3 as function of the red-shift, where the circles denote observational data of super-novae of type Ia (see reference \cite{Riess1}). Only one curve for the coupled model, namely for $\xi_*=0.2$,  was plotted since there is no sensible difference with the $\Lambda$CDM model when $\xi_*\rightarrow0$.  From this figure one concludes that the values of $\mu_0$ for the coupled  model does not differ from the $\Lambda$CDM model for small red-shifts existing only a small depart of the curves for the two models when the red-shift increases.

\begin{figure}
 \begin{center}
 \vskip0.8cm
 \includegraphics[height=6cm,width=8cm]{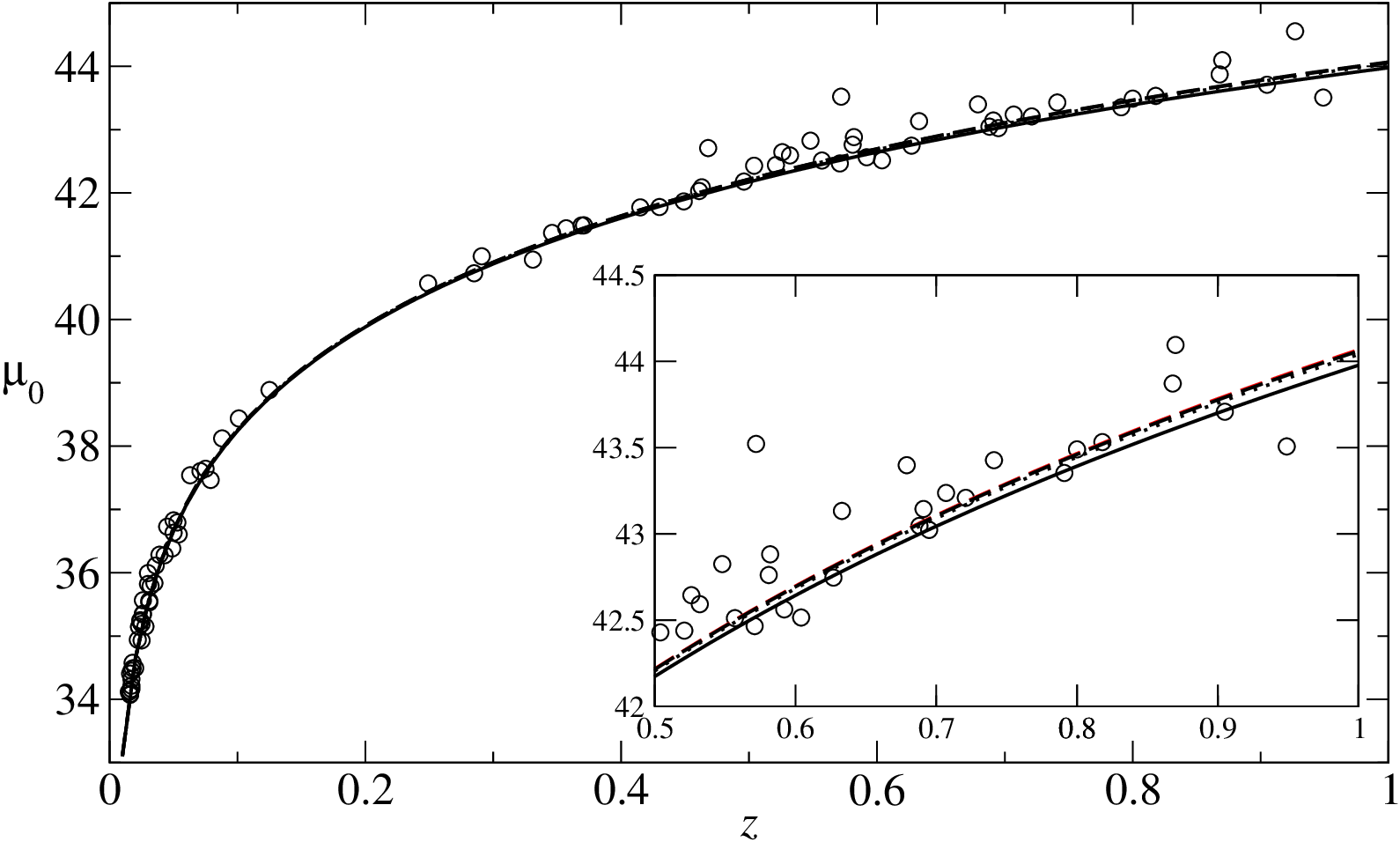}
  \caption{$\mu_0$ as a function of the red-shift $z$. The straight line represents the model with tachyon and matter fields with $\xi_*=0.2$, the dashed line the $\Lambda$CDM model and the circles are observational data of super-novae of type Ia. }
 \end{center}
 \end{figure}

\section{Final remarks}

In the last section the conditions chosen for the scalar field were: (i) $F(0)=\gamma\exp[-\xi\varphi(0)]=1/2$
which corresponds to the value of the gravitational constant presently and (ii) that tachyon field $\varphi$ is
rolling very slowly in the late accelerated period so that $\dot\varphi^2\ll 1$. With these conditions the
action (1) becomes
\be S= \int d^4x\sqrt{-g}\{R/2-V\}+S_m,\label{constcosmol}
\ee{acm}
for the present time. So
one recovers the gravitational action that describes the gravity as it is observed today. Moreover, one has a
potential term with
a cosmological constant behavior which accelerates the actual expansion of the Universe.

{The action (\ref{constcosmol}), which results from the limit of small derivatives of the field
$\varphi$, coincides with the actions analyzed in detail for the models proposed in the references
\cite{Fiziev2, Fiziev1}. Then the limiting situation (\ref{constcosmol}) leads to a constant gravitational
coupling for a constant field $\varphi$. In this limit the present model is similar to the models in the quoted
references which present this limiting propriety, being in accordance with the experimental data. The models
proposed in the works \cite{Fiziev2, Fiziev1} can produce an exponential solution for the scale factor,
characterizing an accelerated expansion. One should hope this same behavior from the field equations of the
present model, once there is a relation between the two models for the limit of small derivatives, and this
really happens. One may infer this kind of solution by observing that for the late time, when the small
derivatives are imposed for the present model, the solutions describe a progressing accelerated Universe while
the energy density of the matter decays and the energy density of the tachyon field increases.}
{Furthermore, the present model can describe the inflationary period with the same function of
coupling and potential which describes the present acceleration by taking the limit of small derivatives, but one should note that in the inflationary case the limit of small derivatives is not taken into account.}

It is also interesting to analyze the evolution equation for the energy density of the coupled tachyon field
$\rho_\v$. By differentiating Friedmann equation (\ref{7}) with respect to time and  taking into account the
acceleration equation (\ref{4}) and the expression for the coupling function (\ref{16}), yields
 \be
 \dot\rho+3H(\rho+p)=-\xi\dot\varphi\rho, \qquad \hbox{hence}\qquad
 \dot\rho_{\varphi}+3H(\rho_{\varphi}+p_{\varphi})=-\xi\dot\varphi\rho,
 \ee{36}
 thanks to $\rho=\rho_m+\rho_{\varphi}$,
$p=p_{\varphi}$ and $\dot\rho_m+3H\rho_m=0.$ From equation (\ref{36})$_2$ one can understand the role played by
the coupling constant $\xi$ on the energy transfer from the tachyon field to the gravitational and matter
fields.

\section{Conclusions}

As was explained before the application of Noether symmetry to the Lagrangian density is a very important tool,
since it guarantees the conservation laws and restricts the possible expressions for the potential density and
for the coupling function of a tachyon field. Here it was shown that according to Noether symmetry only
exponential functions of the tachyon filed are possible expressions for the potential  density and for the
coupling function of non-minimally coupled tachyon fields.

The cosmological solutions found are of two kinds. The first one refers to a non-minimally coupled tachyon field
as the only gravitational source of the early Universe which behaves as an inflaton and leads to an exponential
accelerated expansion which ends in a finite time. In the second one   the gravitational sources of the  Universe
are a non-minimally coupled tachyon field and a pressureless matter field where the tachyon field comports as dark
energy and is the responsible for the decelerated-accelerated transition period of the late Universe. In the latter
 case it was shown that the coupling constant has influence on the density and deceleration parameters and also on
 the luminosity distance, since the coupling constant  is connected with  the energy transfer from the tachyon field
 to the gravitational and matter fields.

\ack
The authors acknowledge the support from CNPq (Brazil).

\section*{References}



\end{document}